\documentclass[aps,pra,10pt,twocolumn,notitlepage]{revtex4-1}
\usepackage{graphicx}
\usepackage{amssymb}
\usepackage{amsfonts}
\usepackage{amsmath}
\usepackage{amsbsy}
\usepackage{natbib}
\usepackage{comment}
\usepackage[colorlinks=true, urlcolor=blue, citecolor=blue,
linkcolor=blue]{hyperref}
\usepackage{color}
\usepackage[utf8]{inputenc}
\usepackage[T1]{fontenc}
\usepackage{amsthm}
\usepackage{mathtools}
\usepackage{lmodern}
\usepackage{graphicx}
\usepackage{mathtools}

\begin{document}

\title[Photoemission via tailored fields]{Photoelectron emission via time and phase-tailored electromagnetic fields}

\author{Jonas W\"{a}tzel, Johannes Hahn and Jamal Berakdar}
\affiliation{Institute for Physics, Martin-Luther-University Halle-Wittenberg, 06099 Halle, Germany}

\begin{abstract}
The energy and the angular distributions of photoelectrons are shown to be tunable by choosing the time and the spatial phase structure of the driving fields. These conclusions are derived from quantum mechanical calculations done within a  single-active electron model for an atomic target subjected to a combination of laser field and a time-asymmetric THz pulse and/or vortex-laser pulse with a spatially modulated phase of the wavefront.
\end{abstract}

\maketitle

\section{Introduction}
The emission process  of electrons upon interaction with photonic fields has not lost its appeal, despite being one of the key experiments triggering the development of quantum mechanics. One of the reasons is probably the critical value of  photoelectron spectroscopy for material and life sciences. While photoelectron spectra gained by weak fields, for example generated in synchrotron facilities \cite{Schmidt2007}, serve to extract the electronic and structural properties of the sample, photoemission generated by tailored short and intense lasers can access dynamical, non-equilibrium states (see further contributions to this special issue and also \cite{Cocker2013,Cocker2016,Danz371}). The goal of this work is somewhat narrow, concentrating on how the photoelectrons can be steered in direction and energy upon modifying the properties of the driving fields. The pulse-generated electrons may serve other purposes as internal electron sources, but the case presented here targets question of  purely fundamental science. \\
The physics of photoemission depends crucially on the pulse duration. For instance, for linearly polarized  long pulses (meaning many optical cycles), the photoelectron spectrum exhibits a forward-backward symmetry. For a pulse duration $\tau_p$  comparable  to $\tau=2\pi/\omega$ ($\omega$ is fundamental frequency), the photoelectron spectrum  disregards  this symmetry, and shows a strong dependence on the carrier-envelope phase \cite{Paulus2001,Paulus2002}. Calculations and  experiments evidence a profound change as the  pulse length varies from $n_p=4$ to $n_p=5$ optical cycles for a sin$^2$ temporal envelope function.\\
A prominent feature of photoelectrons produced via above-threshold ionization (ATI) concerns  the high energy part of the spectrum showing a plateau-like structure, first observed for  40~fs pulses~\cite{Paulus1994}. Furthermore,  ATI peaks at energies
in the range of around 6-10 times the ponderomotive potential $U_p$, each with a distinct
intensity threshold for their onset~\cite{Hertlein1997}. The re-scattering model of Corkum~\cite{Corkum1993} offers
a qualitative explanation in terms of three steps: electron tunneling into continuum caused by the laser field, an electron wave packet bounce-back when the laser field changes sign, and rescattering on the residual ion.\\
Here, we investigate the effect on the photoelectron spectra when the driving field is superimposed with a short time-asymmetric pulse. The former is either an unstructured, conventional laser pulse or an optical vortex carrying orbital angular momentum. Photoelectrons attain from the time-asymmetric pulse linear momentum  \cite{MOSKALENKO20171} and from the optical vortex angular momentum \cite{schmiegelow2016transfer,afanasev2018experimental,DeNinno2020,picon2010photoionization, seipt2016two, watzel2016discerning}. Quantum dynamic simulations for the whole process are done by numerically solving the time-dependent Schr{\"o}dinger equation ~\cite{Wassaf2003b,Tong2007}.\\
%
As dictated by Maxwell's equations, the time-integral over the electromagnetic field vanishes for propagating pulses. However, one half-cycle of the pulse may be much shorter and stronger than the other half with opposite polarity \cite{Jones1993,Wesdorp2001,Mestayer2007}.
Such pulses are  widely produced in the THz-regime and referred to as half-cycle pulses (HCP) since the half-cycle with the strong peak field affects the electron dynamics most. For a review we refer to \cite{MOSKALENKO20171}. Under certain conditions, the HCP interaction with atoms can be described as a momentum transfer $\pmb{q}=\pmb{e}\int _{-\infty}^\infty F(t)dt$, with $\hat{e}$ being the polarization vector of the short pulse. In this limit, the theoretical description is significantly simplified ~\cite{Krstic1994,Dimitrovski2004, MatosPhD, Dimitrovski2006, Matos2007, Grozdanov2009}, yet the physics is still very rich basically because of the very wide bandwidth of the short pulse (further caveats are discussed in \cite{MOSKALENKO20171}).
The simulations presented in Sec. III, after a brief theory description in Sec. II, are for an argon atom irradiated by a combination of a linearly polarized few-cycle field with a unidirectional HCP. In a second step, we replace the linearly polarized field with a circularly polarized optical vortex carrying total angular momentum of $2\hbar$. The paper ends with brief conclusions and outlook.\\
Throughout the paper, we use atomic units unless indicated otherwise.


\section{Model and method}
Rare gas atoms are common targets for ATI experiments~\cite{Paulus1994,Nandor1999}.  To model photoelectron spectra from multielectron atoms we rely
on a single active electron (SAE) approximation. For  $x$-linearly polarized pulses, we
use  the Kohn-Sham-type (KS) potential for the $3p$ valence electron:
\begin{equation}
v(x)=-\frac1{\sqrt{c+x^2}}.
\end{equation}
A crucial point is the correct asymptotic behavior of the potential $v(x)\sim1/x$ for
$x\rightarrow\infty$ implying an infinite number of bound states. The
properties of the potential were studied in details in~\cite{Su1991} and were used
to model  ATI of rare gas atoms~\cite{Paulus1994,Wassaf2003b}. The numerical
value of the constant $c=1.41$ is selected to match the energy of the $3p$ electron (the
negative of the ionization potential) in Ar ($\epsilon_0=-0.58$ a.u.). The numerical solution
of the Schr{\"o}dinger equation is performed on an equidistant grid in space and
time with a length $L=2970$ (step $\Delta x=0.1$\,a.u.) and a time step $\Delta t=0.04$ a.u. yielding the time-dependent wave function $\psi(x,t)$. In the 1D case, the photoelectron spectrum for finding the electron with the energy $\widetilde{E}_i > 0$ is given by \cite{javanainen1988numerical}
\begin{equation}
P(\widetilde{E}_i)=\frac{\langle\psi_i|\psi(t)\rangle}{E_{i+1}-E_{i-1}} + \frac{\langle\psi_{i+1}|\psi(t)\rangle}{E_{i+2}-E_{i}}
\end{equation}
which is an average over the populations of the even eigenstates $\psi_i$ and odd eigenstates $\psi_{i+1}$ (with Eigenenergies $E_i$ and $E_{i+1}$) of the atom. The average (kinetic) energy $\widetilde{E}_i=(E_{i-1} +E_{i} +E_{i+1} +E_{i+2})/4$. As emphasized in Ref.\,\onlinecite{javanainen1988numerical}, the purpose is the canceling of vast oscillations between the populations of  states with different parity. \\
When applying an optical vortex, the problem is three-dimensional that we formulate in spherical coordinates $\{r,\vartheta,\varphi\}$. The radial single-particle potential is given by \cite{muller1999numerical}
\begin{equation}
V(r) =-(1 + 5.4r + 11.6e^{-3.682r})/r,
\end{equation}
which was employed in similar problems \cite{toma2002calculation}. The interaction Hamiltonian, which describes the action of the laser fields, follows from minimal coupling as
\begin{equation}
\hat{H}_{\rm int}(t)=\pmb{A}(\pmb{r},t)\cdot\hat{\pmb{p}} + \frac{1}{2}\pmb{A}^2(\pmb{r},t),
\end{equation}
where $\pmb{A}(\pmb{r},t)$ is the combined vector potential of all present fields.
For the numerical solution of the three-dimensional Schr{\"o}dinger equation we  expand the wave function in spherical harmonics $Y_{\ell,m}(\Omega_{\pmb{r}})$, i.e., $\Psi(\pmb{r},t)=\sum_{\ell,m}R_{\ell,m}(r)Y_{\ell,m}(\Omega_{\pmb{r}})$. Applying the matrix iteration algorithm \cite{nurhuda1999numerical, grum2010ionization}, the angular channels corresponding to the quantum numbers $\ell$ and $m$ are time-propagated until all fields are off.
The photoionization probability (differential cross section, DCS) is calculated from the propagated wave function by projecting onto the scattering states $|\varphi_{\pmb{k}}^{(-)}\rangle$:
\begin{equation}
w(\pmb{k})=\left|\langle\varphi_{\pmb{k}}^{(-)}|\Psi(t>T_{\rm obs})\rangle\right|^2.
\label{eq:DCS}
\end{equation}
Here, $\pmb{k}$ is the asymptotic momentum of the photoelectron with the kinetic energy $E_k=k^2/2$, and $T_{\rm obs}$ is an observation time during which all external light fields have vanished. The scattering states $|\varphi_{\pmb{k}}^{(-)}\rangle$ solve the respective time-independent Schr\"odinger equations for positive energies $E_k$.\\
For three-dimensional problems, $w(\pmb{k})$ depends on the asymptotic direction $\Omega_{\pmb{k}}$ of the photoelectron, where $\Omega_{\pmb{k}}=\left\{\vartheta_{\pmb{k}},\varphi_{\pmb{k}}\right\}$ is the solid angle in  $\pmb{k}$-space. The total cross section is given by
\begin{equation}
\sigma(E_k)=\int{\rm d}\Omega_{\pmb{k}}\,w(\pmb{k}).
\end{equation}

\section{Results}
\subsection{Single HCP, effect of the shape}

\begin{figure}
\centering
\includegraphics[width=0.9\columnwidth]{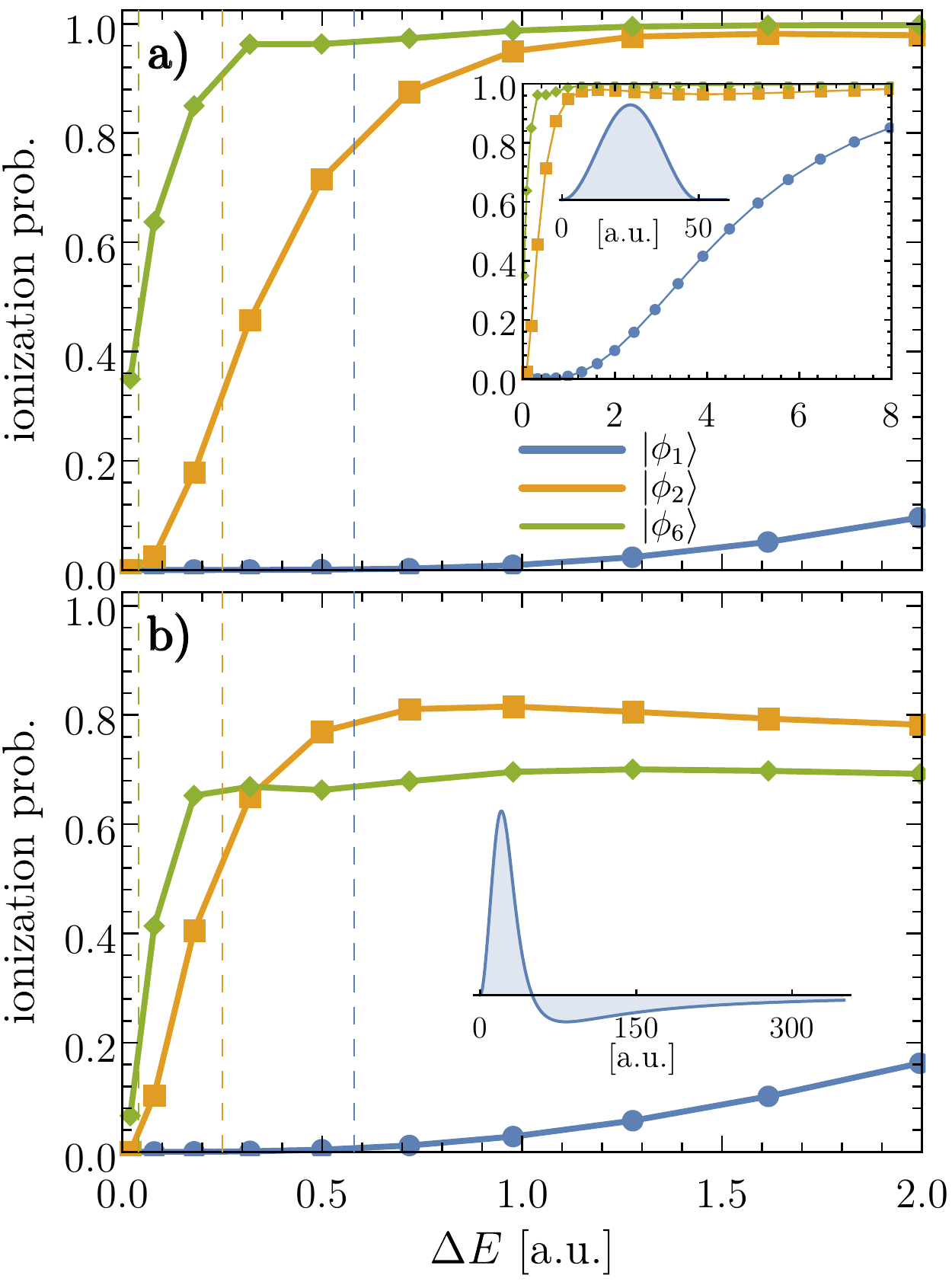}
\caption{Ionization probability for different initial states by applying HCPs with a temporal function described by $\mathcal{F}_{\rm HCP}^{(1)}(t)$ or $\mathcal{F}_{\rm HCP}^{(2)}(t)$ HCP (Eqs.~\ref{eq:hcp1},~\ref{eq:hcp2}) in  dependence on the energy shift $\Delta E_k$ ($\Delta E_k$ refers to the first half cycle). Vertical lines denote the estimated ionization thresholds.\label{fig:hcp}}
\end{figure}

Before considering a more involved case of the combined action of few- and half-cycle
pulses, let us focus on the action of the latter. For atomic systems, the photoionization
probabilities were analytically obtained for single HCP~\cite{Krstic1994} and for a
sequence of HCPs~\cite{Dimitrovski2004} by using the
Magnus expansion~\cite{Magnus1954,Pechukas1966}, which provides to first-order useful results for short pulses of arbitrary strength (for more details on the property of this expansion in connection with this work we refer to \cite{MOSKALENKO20171}). The
time evolution operator is written in to first order as
\begin{equation}
U(0,\tau_p)=\exp\left(-i\pmb{r}\cdot\int_0^{\tau_p}\pmb{E}(t)\,{\rm d}t\right).\label{eq:uhcp}
\end{equation}
We assume here that the laser field $\pmb{E}(t)=\pmb{e}F(t)$ is switched on at the time $t=0$, while
the pulse duration is $\tau_p$. The evolution  depends on the momentum $\pmb{q}=\pmb{e}\int_0^{\tau_p} \vec{F}(t)\,{\rm d}t$ transferred to the system and is  independent of the pulse shape. Hence, all pulses with the same momentum transfer $q$ are equivalent to the delta-function pulse shape $F(t)=\delta(t)$ for which the approximation is,
actually, the correct result. It is possible to analytically
treat a full-cycle pulse by modeling   it by two $\delta$-pulses separated by a
half-cycle time $\tau_p$~\cite{Dimitrovski2004}, the long tail of the actual  HCP cannot be described within this
approach (which is necessary  when its duration becomes comparable with internal time-scale of the system). In  the present study, the (first order) Magnus approximation is insufficient. Thus,
we have to resort to numerics. At first, we compare the probability to ionize the system by
using HCP of two different temporal shapes:
\begin{equation}
\mathcal{F}_{\rm HCP}^{(1)}(t)=E_0\sin^2\left(\frac{\pi t}{\tau_p}\right),
\label{eq:hcp1}
\end{equation}
or
\begin{equation}
\begin{split}
\mathcal{F}_{\rm HCP}^{(2)}(t)=&-E_0\exp\left(-\frac{t^2}{2\kappa^2a^2}\right)
\frac{bt^2}{b^2+(t/\kappa)^4}\\
&\times\sin\left(\frac{2\pi}{1+(t/\kappa)^{1/4}}\right).
\end{split}
\label{eq:hcp2}
\end{equation}
The second complicated form of the pulse was selected to fulfill the condition
$\int_{-\infty}^{\infty}\mathcal{F}_{\rm HCP}^{(2)}(t) {\rm d}t=0$ and to have a possibility to change the
ratio of amplitudes of the half cycles with opposite polarity. In Fig.~\ref{fig:hcp}, the ionization probability is plotted as a
function of the transferred mean energy $\Delta E=q^2/2$ for the two types of HCP pulses.  Upon the momentum transfer $q$, the kinetic energy of a photoelectron with asymptotic momentum $k$ is shifted by
\begin{equation}
\Delta E_k=kq+\frac12q^2.
\label{eq:eng-balance}
\end{equation}
Because of the specific form of the time-evolution operator (Eq.~\ref{eq:uhcp}), the ionization
probability reflects the momentum distribution of the ionized state~\cite{Wetzels2003}. The vertical dashed lines
present the location of the state-specific ionization thresholds, i.e., $\Delta E_k=-\epsilon_i$.
For the simulations, we used HCPs with a duration 50 a.u. (first half cycle), which were
applied to the system in the ground and excited states. Although the electron orbiting
times  $\tau_i$ in these states are considerably shorter than the pulse duration, one
observes satisfactory agreement between the position of the steep increase of the ionization probability and the
threshold energies. For instance, the probability of ionization at  threshold is approximately 32\% for the second state. For the 6th state, it is already 42\%. In contrast, for the ground state $\tau_p\gg\tau_1$, so that the Magnus approximation is insufficient, which is highlighted by the negligible ionization probability at the threshold energy.\\
The HCP application with zero momentum transfer  (Fig.~\ref{fig:hcp}, right panel) does
not qualitatively change the picture of the ionization observed for the ideal HCP. There are, however, critical quantitative differences. First, the system can never be fully ionized due to recombination. Second, different states reach different saturated levels of ionization. Further, at much larger values of the energy transfer, the ionization slowly drops down. Such behavior is anticipated for higher energy states or shorter pulses. In Ref.~\onlinecite{Wesdorp2001}, it was experimentally shown that the long tails of realistic HCPs start to be significant only when the duration of the whole pulse is in the regime of impulsive kick, i.e., when the orbiting time of the electron is longer. In our  one-dimensional calculations, a sizeable effect of the long tail of HCP appears for much longer pulse lengths. Further, we note the clear difference to ionization with
two alternating $\delta$-pulses \emph{leading to the full ionization of the system}
at sufficiently large momentum transfer~\cite{Grozdanov2009}.\\
Second finding is the observation of an ionization onset at lower fields by using $\mathcal{F}_{\rm HCP}^{(2)}(t)$ instead of $\mathcal{F}_{\rm HCP}^{(1)}(t)$. For example, at the threshold energy $E_2$, one achieves an ionization probability of almost 10\% higher for the second state (43\% to 52\%). This  effect was also observed
experimentally \cite{Wesdorp2001}.

\subsection{Off-resonant excitation and ionization with HCP}
\label{sec:B}

\begin{figure}
\centering
\includegraphics[width=0.9\columnwidth]{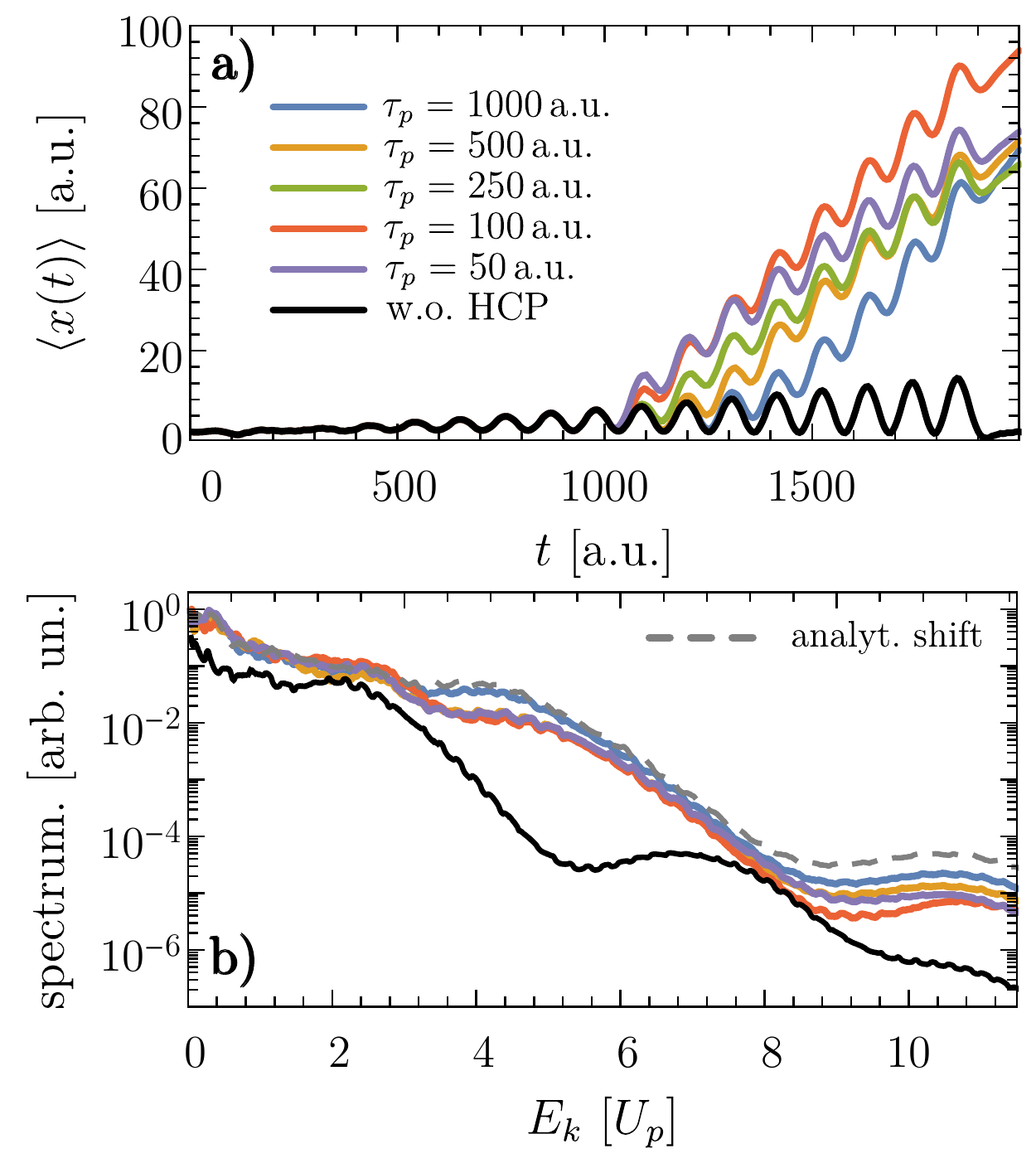}
\caption{The time-dependent expectation value of the position (left panel) and photoelectron spectrum (right panel) for HCPs of different durations while the momentum transfer of 0.42~a.u. kept fixed. The HCPs are applied to the system at $t=1000$~a.u.. Colored curves correspond to the application of IR laser field with HCP, and the black curve to the case when only the HCP is present. Analytical energy expected shift is indicated by the gray, dashed curve in panel b). }
  \label{fig:ati2}
\end{figure}

The application of short electromagnetic pulses of high intensity to an atomic system leads
among other effects to  above threshold ionization (ATI). Wassaf \emph{et
al.} performed detailed studies of the present one-dimensional model system~\cite{Wassaf2003a,Wassaf2003b}. Their results closely mimicked experimental observations and explained the resonance-like enhancements in ATI
spectra. The drawn conclusions on the origin of the effect can be considered complementary to another explanation based on the "channel closing"-hypothesis~\cite{Popruzhenko2002,Borca2002}. In the present work, we  reproduced
the aforementioned spectra, albeit with a different numerical method.
\begin{figure}
\center{\includegraphics[width=0.9\columnwidth]{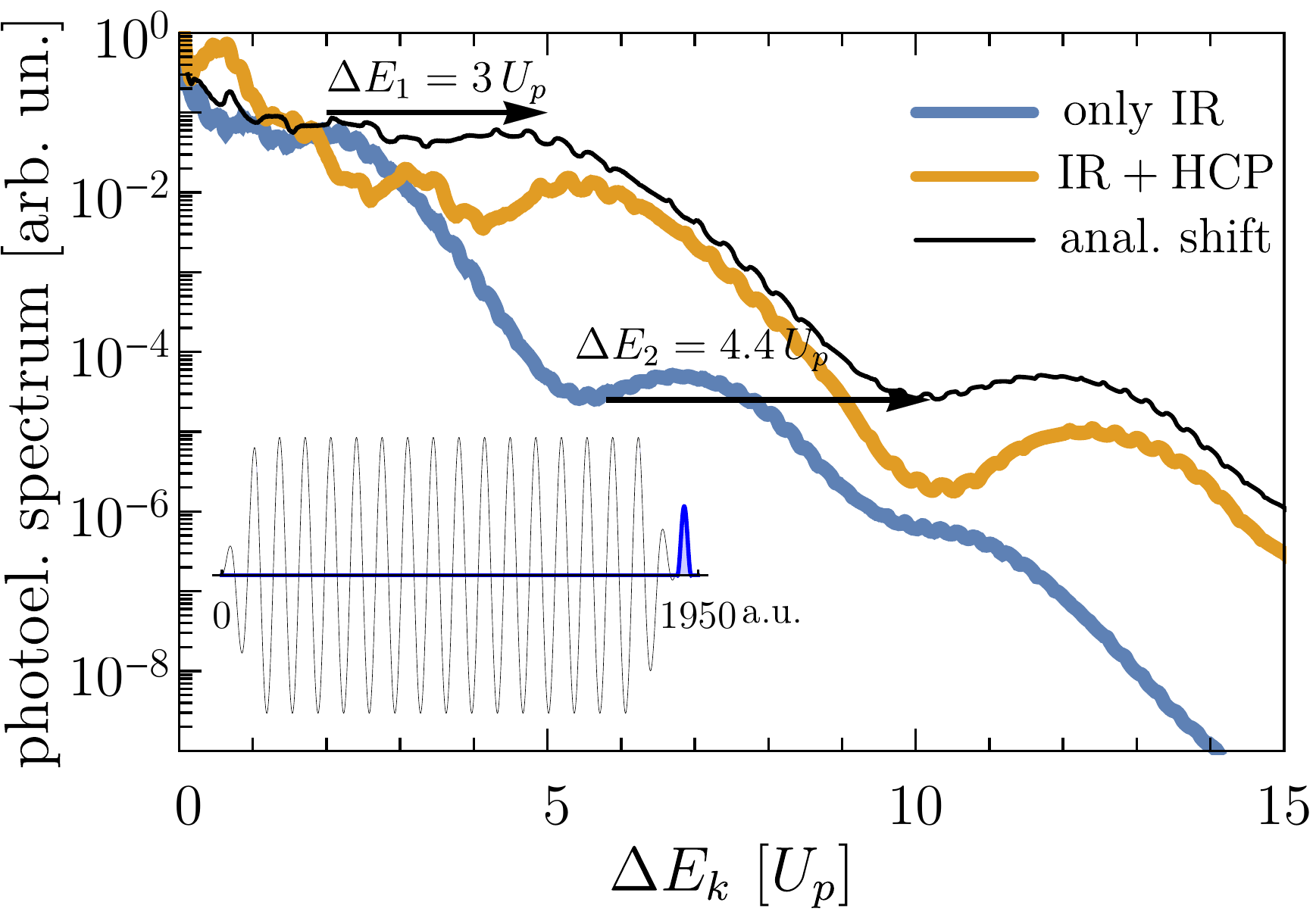}}
\caption{Photoelectron spectrum as a function of the energy of ejected electrons measured
  in units of the ponderomotive potential ($U_p$). The system is excited with a trapezoidal
  pulse (shown in the inset) with an intensity of $1.12\cdot10^{14}$~W~cm$^2$ and
  $\hbar\omega=1.57$~eV. We compare the ATI spectrum obtained without and with the presence of a HCP
  with a duration of 50 a.u. applied at the end of the first pulse. The solid black
  line denotes the hypothetical ATI spectrum obtained by applying the analytical energy shift (cf.\,Eq.~\ref{eq:eng-balance}).
\label{fig:ati1}}
\end{figure} Furthermore, we
study the effects of an additional HCP.
The atom is excited  with an electromagnetic pulse of 1960~a.u. duration and frequency
of 0.0577~a.u., the energy of the photon is typical for Ti:sapphire laser. The pulse comprises 18
optical cycles and has an intensity of $1.12\cdot10^{14}$~W~cm$^2$ (Fig.~\ref{fig:ati1},
inset). It should be noted that a straightforward application of
these equations leads to oscillations in the ATI spectra which can be removed by
averaging. We  present only averaged spectra noting that several components of these oscillations are discussed in detail in Ref.~\cite{Wassaf2003a}. The distance between major peaks corresponds to the energy of one
photon, while subpeaks signify processes involving excited states.\\
In our first numerical experiment (Fig.~\ref{fig:ati1}), we study the influence of the HCP, which is
applied at the end of the few-cycle pulse, on the ATI spectrum. The HCP has a
sin$^2$ shape with 50~a.u. duration and $E_0=2.24\cdot10^{-2}$~a.u. maximal electric field
strength. The HCP is applied at the moment $t=1950$~a.u. so that we find a small interval of time when both pulses intersect. The energy transferred to the system by the HCP ($\Delta E=0.16$~a.u.) is insufficient to ionize the
system. However, its influence is prominent. It provides not only the expected energy shift (analytical treatment of this effect is depicted by the black line) of the spectrum but also substantially reduces the electron yield and introduces additional peaks
(Fig.~\ref{fig:ati1}, cf. orange curve). The first peak with the energy of $0.65U_p$ originates from  stripping off the electron population from the higher excited states due to the HCP action. This statement is supported by the energy balance ($U_p=0.24$~a.u., $q=0.56$~a.u., $\Delta E=0.65U_p$). However, the origin of the peak at $3U_p$ and the smaller
photoelectron yield cannot be inferred from Fig.\,\ref{fig:ati1} alone. Hence, we performed
a series of calculations where we apply  HCPs with different durations ranging from $\tau_p=50$~a.u. to
$\tau_p=1000$~a.u. and a fixed momentum transfer of 0.42~a.u.. We do so while the atom is subjected to  the longer laser pulse (Fig.~\ref{fig:ati2}).
The ATI spectrum depends strongly  on the duration of the pulse. Longer
pulses have a weaker influence on the electron yield, as shown in Fig.~\ref{fig:ati2} (blue and orange curves). Shorter pulses,  reduce significantly  the emission of  electrons with larger kinetic energies. This effect is more pronounced in the vicinity of the resonant enhancements, in line with the classical description of  electron rescattering during  ATI, as was originally suggested by Paulus \emph{et al.} for two-color ATI~\cite{Paulus1995}.

\subsection{Combination of optical vortex and HCP}

\begin{figure}[t!]
\centering
\includegraphics[width=0.85\columnwidth]{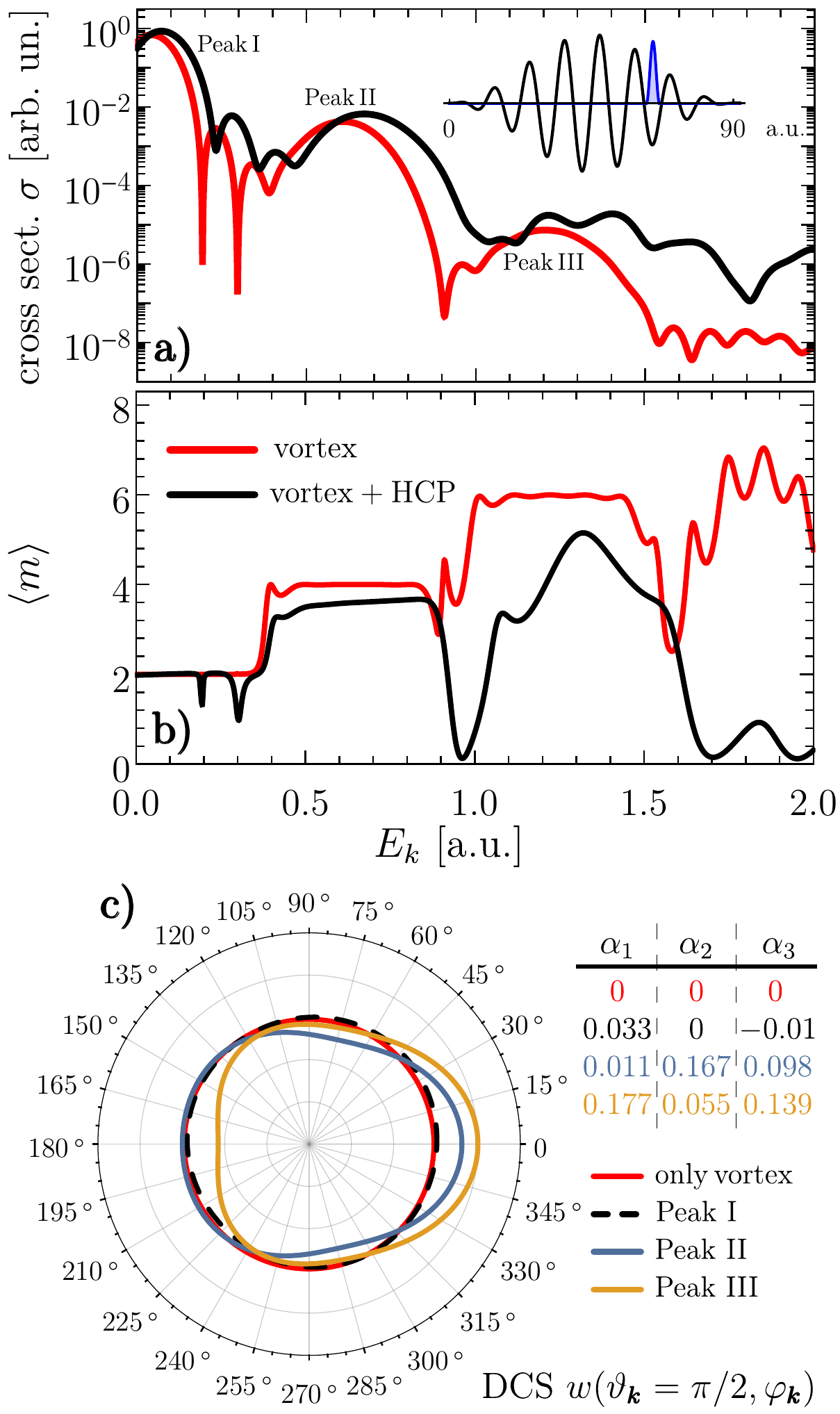}
\caption{Ionization of the  Argon atom using a combination of the optical vortex and the attosecond half-cycle pulse (shown in the inset of panel b). (a) The photoemission cross-section for using only a vortex and the combination. (b) Angular momentum attained by the photoelectron. (c) Angular distribution of the photoelectrons in the polarization plane and the $\beta$ parameters \cite{Schmidt2007} (see main text) for the different major peaks in the photoemission spectrum. The black curve corresponds to applying only the vortex. For  clarity, all DCS are normalized to the same total cross section $\sigma$.}
  \label{fig:ov}
\end{figure}

For the optical vortex propagating along the $z$-axis, the helicity $\sigma$ and carried photonic orbital angular momentum $m$ have the same sign (total angular momentum is $(m+\sigma)\hbar$). Such a structured field is called a parallel class vortex and can be described by a transverse vector potential, so that $\pmb{\nabla}\cdot\pmb{A}(\pmb{r},t)=0$ is valid to an very good approximation \cite{watzel2020electrons}. The  atom is located at the optical axis. In such a case, the vector potential is given in cylindrical coordinates $\pmb{r}=\{\rho,\varphi,z\}$:
\begin{equation}
\pmb{A}(\pmb{r},t)=A_0\pmb{e}\left(\frac{\rho}{w_0}\right)^{|m|}e^{im\varphi}\Omega(t)e^{-i\omega t} + c.c.,
\end{equation}
where $A_0$ is the amplitude, $w_0$ is the beam waist, $\Omega(t)$ is the temporal envelope and $\pmb{e}_{\sigma}=(\pmb{e}_{\rho}+i\sigma\pmb{e}_{\varphi})\exp(i\sigma\varphi)$ is the circular polarization vector. In the following, we consider the argon atom to be irradiated by the vortex with $\sigma=+1$, $m=+1$, and a photon energy of $\hbar\omega=14.96$\,eV (0.55\,a.u.). The absorption of one vortex photon is not sufficient to ionize from the ground state, but from higher unoccupied states if they were photo-excited by another pulse. Multi-photon processes are captured by the numerical propagation and result in photoelectron spectra with major peaks separated by $\hbar\omega$.\\
The results shown in Fig.\,\ref{fig:ov} correspond to a $\sin^2$ envelope with a pulse length of 8 optical cycles. The amplitude belongs to a peak electric field of $E_0=5$\,a.u. at the donut rim (waist $w_0=50$\,nm) of the vortex. However, since the argon atom is located in the singularity (where $\pmb{A}(\rho=0,t)=0$), the valence shell electron is exposed to a relatively weak electric field. The red curve in Fig.\,\ref{fig:ov} corresponds to the case when only the vortex field is applied, while the black curve belongs a combination of a vortex and HCP pulses. In panel b), we present the $z$-projection of the angular momentum $\langle m\rangle=\langle\Psi(t)|\hat{L}_z|\Psi(t)\rangle/\langle\Psi(t)|\Psi(t)\rangle$ attained by the photoelectron in the dependence on the kinetic energy $E_k$. With every absorption of a vortex photon (in absence of the HCP), an angular momentum exchange of $2\hbar$ occurs so that at every major peak in the spectrum, see  Fig.\,\ref{fig:ov}a), $\langle m\rangle$ increases by two. The third peak around $E_k=1.8$\,a.u. is not very well developed, which can be explained by the low local field amplitude around the singularity.
Now let us switch on an attosecond HCP \cite{wu2012giant, xu2018terawatt} described by $\mathcal{F}_{\rm HCP}^{(1)}(t)$ with a duration of $\tau_p=100$\,as (4.13\,a.u.) and $E_0=0.05$\,a.u. yielding a momentum kick $q=0.103$\,a.u. ($\Delta E=0.005$\,a.u.). The temporal profiles of both pulses are shown in the inset in Fig.\,\ref{fig:ov}a). Relative to the optical vortex (which propagates in $z$-direction), the HCP is directed along the $x$-axis, meaning it acts in the same plane as the vortex vector potential $\pmb{A}(\pmb{r},t)$. Similar to Sec.\,\ref{sec:B}, the HCP has a clear impact on the photoelectron spectrum and $\langle m\rangle$.
On the one hand, the maxima of the cross section are shifted, which can be attributed to the attosecond HCP momentum kick and follows roughly Eq.\,\eqref{eq:eng-balance}. On the other hand, the photoelectron yield is visibly increased. Further, the attained angular momentum is clearly damped due to the action of the HCP, so that the "stair"-like behavior ($\langle m\rangle$ increases by two for every major peak) does not apply to the combined irradiation of both laser fields. Strong effects can be observed for higher energies $E_k>1.5$\,a.u., where the effect of the vortex (relative to the HCP) is virtually non-existent.
Therefore, also $\langle m\rangle$ is practically zero. \\
The differential cross section (DCS) has an angular dependence of the two-laser excitation/ionization process. As both electromagnetic fields act in the $x-y$ plane, we  study the azimuthal variation of the DCS at $\vartheta_{\pmb{k}}=\pi/2$, which is shown  by Fig.\,\ref{fig:ov}c). Without the HCP (only the vortex field is present, red curve), the DCS is invariant with respect the azimuthal angle variation which stems from the cylindrical symmetry of the vortex field. In the presence of the HCP, however, the forward-backward symmetry of the photoionization process is broken and the degree of the asymmetry depends on the kinetic energy of the photoelectron. Here, the depicted DCS curves are averaged over the widths of the major peaks in the photoelectron spectrum, shown in Fig.\,\ref{fig:ov}a). Observing the photoelectron at a higher (major) peak in the cross section, increases the degree of asymmetry. For a more clear quantification of the symmetry break, we can fit the (peak-averaged) DCS curves to
\begin{equation}
\bar{w}_{p}(\vartheta_{\pmb{k}}=\pi/2,\varphi_{\pmb{k}})\propto1 + \sum_{n}^3\alpha_nP_n(\cos(\varphi_{\pmb{k}})),
\label{eq:beta}
\end{equation}
where $p$ is the index of the observed major peak, $P_n(x)$ are the Legendre polynomials of degree $n$, and $\alpha_n$ can be interpreted as the asymmetry parameter for the azimuthal dependency. We note that in comparison to the conventional introduction of asymmetry parameter $\beta$ \cite{taylor1977photoelectron, hemmers1997beyond, dixon2005recoil, child2008limits} (which describes the full solid angle dependency of the DCS), this is a very simple model whose sole purpose is the quantification of the DCS deformation in the azimuthal plane due to the action of the HCP. While $\alpha_2$ reflects symmetry between $\varphi_{\pmb{k}}$ and $\varphi_{\pmb{k}}-\pi$, $\alpha_1$ and $\alpha_3$ quantify asymmetry. We note that in most of the existing literature, the expansion in Eq.\,\eqref{eq:beta} refers to the polar angle $\vartheta_{\pmb{k}}$ for $z$-linearly polarized fields. Here, however, both polarization vector point in the azimuthal plane, so that an expansion in the dependence on $\varphi_{\pmb{k}}$ is reasonable.\\
Applying the vortex field only (without the HCP) yields a cylindrically symmetric DCS for all three ionization peaks. Therefore, all $\alpha_n$ are virtually zero. Adding now the HCP causes  an asymmetry in the photoelectron's momentum distribution. Physically, the valence shell electron is now preferentially ejected into the HCP polarization direction ($x>0$). Moreover, this effect is even more pronounced when observing the photoelectron at higher kinetic energies $E_k$. This is highlighted by the increased $\alpha$ parameter with odd $n$. As indicated by Fig.\,\ref{fig:ov}b), the presence of the HCP is -- at least for the first two peaks -- not obstructive for the angular momentum transfer to the electron, i.e., $\langle m\rangle$ shows the same trends. Therefore, the application of an HCP has the advantage that we can steer the emission direction of a photoelectron with encapsulated angular momentum.
\section{Conclusions}
This numerical study shows that the photoelectron spectra produced in an above-threshold-ionization of an atom in an intense laser field can be modified by applications of time-asymmetric pulse that shifts the linear momentum of the electron. The amount of this shift is set by the duration and the peak electron field of the applied pulse. If the laser field is an optical vortex carrying orbital angular momentum, the photoelectrons may attain an electronic angular shift that we can quantify by the expectation value of the orbital angular momentum along with the propagation directions. This orbital angular momentum can be quenched or maintained by applying appropriate time asymmetric pulses.

\begin{acknowledgments}
The authors thank Yaroslav Pavlyukh for enlightening and fruitful discussions. This study is supported by funded by the Deutsche Forschungsgemeinschaft (DFG) under  SPP1840,  WA4352/2-1, and SFB TRR227.
\end{acknowledgments}

%

\end{document}